\documentclass[fleqn,twoside]{article}
\usepackage{espcrc2}

\usepackage{graphicx}
\usepackage[figuresright]{rotating}

\newcommand{\AmS}{{\protect\the\textfont2
  A\kern-.1667em\lower.5ex\hbox{M}\kern-.125emS}}

\newcommand{\lsim}{ \mathop{}_{\textstyle \sim}^{\textstyle <} }

\title{EDMs in  SUSY GUTs}

\author{Junji Hisano \address[MCSD]{ICRR, University of Tokyo,  5-1-5 Kashiwa-no-Ha
Kashiwa City, 277-8582, Japan}%
}
       
\begin{document}

\begin{abstract}
Hadronic and leptonic EDMs in SUSY GUTs are reviewed in this article.  
\vspace{1pc}
\end{abstract}

\maketitle

\section{Introduction}

SUSY GUTs may predict rich flavor violation, and the signature may
be observable in the low-energy flavor physics. The SUSY breaking
terms in the minimal supersymmetric standard model (MSSM) are
sensitive to physics beyond the MSSM. When the origin of the SUSY
breaking terms in the MSSM comes from dynamics around or above the
GUT scale, the interactions in the SUSY GUTs may bring down the
signature on the SUSY breaking terms.

In the MSSM the sizable flavor-violating SUSY breaking (FVSB) terms for the
left-handed squarks are induced by the large top-quark Yukawa
coupling, and those for the left-handed sleptons may be also generated
by the neutrino Yukawa interaction in the SUSY seesaw mechanism. In
the SUSY GUTs the FVSB terms for those SU(5)
partners, the right-handed squarks and sleptons, are generated. This
gives a chance to probe the interactions at the GUT scale by the
low-energy flavor-changing processes, such as the $K^0$--$\overline{K}^0$
mixing, the $B$ physics, and the lepton flavor violation.

The hadronic and leptonic electric dipole moments (EDMs) are sensitive
to beyond the SM since the SM contributions are suppressed. In the
MSSM the one-loop diagrams of SUSY particles may contribute to
them. The EDMs are flavor-conserving observables, however, they are
sensitive to the FVSB in the internal lines of the loop
diagrams. Especially, when both left-handed and right-handed squarks
or sleptons have the FVSB terms with the CP phases,
the EDMs are enhanced by the heavier fermion mass. Thus, the EDMs are
good probes for the SUSY GUTs.

In this paper we review the prediction of the hadronic and leptonic
EDMs in the SUSY GUTs, and the implication to the other observables in
the models.

\section{Hadronic EDMs}

When partons in nucleon have CP violating interactions, they
contribute to the hadronic EDMs, such as $^{199}$Hg and neutron
EDMs. The current experimental bounds on neutron and $^{199}$Hg atom
EDMs are $ |d_{n}| < 6.3 \times 10^{-26} e\, cm$ and $ |d_{\rm Hg}| <
1.9\times 10^{-28} e\, cm$, respectively (90\%C.L.) \cite{edmexp}.
The CP violation in the strong interaction of the light quarks is
dictated by the QCD theta term and the quark CEDMs,
${d}^C_q~(q=u,d,s)$, up to the dimension five terms. When the
Peccei-Quinn symmetry is imposed,  the QCD theta
parameter is suppressed.

The neutron and $^{199}$Hg atom EDMs are evaluated as $ d_{\rm
Hg}=-8.7\times 10^{-3}\times e(d_u^C-d_d^C+0.005 d_s^C)$ and $d_n =
-1.6 \times e (d_u^C+0.81\times d_d^C+0.16\times d_s^C)$
\cite{Hisano:2004tf}.  The
contributions from the strange quark CEDM are included. From the SU(3)
octet baryon mass splitting and the sigma term in the $N$--$\pi$
scattering, it is found that the strange quark component in nucleon is
not negligible compared with other light quarks
\cite{Zhitnitsky:1996ng}.

Since the $^{199}$Hg atom is diamagnetic, the EDM is sensitive to
CP-violating nuclear force induced by the meson exchange, which
generates T-odd EM potential parametrized by the Shiff moment.  The
latest evaluation of the Shiff moment \cite{Dmitriev:2003kb} reveals
that the the Shiff moment is sensitive to the isovector channel of the
meson exchange. Thus, the strange quark CEDM contribution is
suppressed by $\pi^0$--$\eta^0$ mixing. On the other hand, the neutron
EDM is generated by nucleon-meson one-loop diagrams. The strange quark
CEDM contribution to the neutron EDM is sizable compared with other
light quarks. Here, the local counter term contribution to the neutron
EDM is not included since it is not constrained by other
observables. The theoretical evaluation still has large theoretical
uncertainties.

From the experimental bound on the neutron ($^{199}$Hg) EDM, the
CEDMs are bounded as $e|{d}^C_u|<3.9(2.2)\times 10^{-26}\;e\;cm$,
$e|{d}^C_d|<4.8(2.2)\times 10^{-26}\;e\;cm$, and $e|{d}^C_s|<2.4(44)
\times 10^{-25}\;e\;cm$,
assuming that the accidental cancellation among the CEDMs does not
suppress the EDMs.  Now the measurement of the deuteron EDM using the
storage ring is proposed, and it is argued that the sensitivity may
reach $d_D\sim 10^{-27}~e~cm$ \cite{Semertzidis:2003iq}. This corresponds
to $d_u^C\sim d_d^C\sim10^{-28}~e~cm$ and $d_u^C\sim
d_d^C\sim10^{-26}~e~cm$ \cite{Hisano:2004tf}.  Thus, the further
improvement may be possible.

As mentioned in Introduction, the hadronic EDMs are predicted by the
FVSB in the SUSY GUTs. While the neutrino Yukawa coupling generates
the right-handed sdown mixings, the FVSB induced by the top quark
Yukawa coupling in cooperation of the CKM mixing is expected to
dominate in the left-handed sdown mixings. Thus, we may probe the
neutrino Yukawa coupling by using the hadronic EDMs
\cite{Hisano:2004pw}.

\begin{figure}[htb]
\begin{center}
\includegraphics[width=18pc]{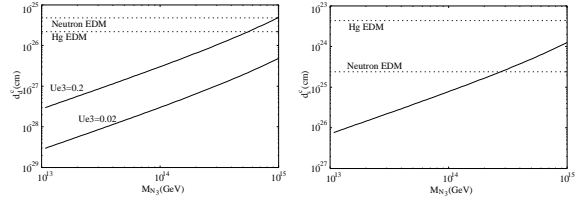} 
\caption{
CEDMs for the strange quark in (a) and for the down quark in (b) as
functions of the right-handed tau neutrino mass, $M_{N_3}$. Here,
$U$ is  the MNS matrix. See
Ref.~\cite{Hisano:2004pw} for the input parameters.}
\label{fig:edm1}
\end{center}
\end{figure}

In Fig.~\ref{fig:edm1} the down and strange quark CEDMs in the SUSY
SU(5) GUT with the right-handed neutrinos are presented
\cite{Hisano:2004pw}. The right-handed sdown mixing between the second
and third generations, depend on $U_{\mu3}$ with $U$ the MNS matrix
for the neutrino mixing, generates $d_s^C$. It is found from this
figure that the right-handed tau neutrino mass larger than about
$10^{14}$ GeV are excluded from the bound on $d_s^C$. On the other
hand, $d_d^C$ is sensitive to $U_{e3}$ via the correction to the the
right-handed sdown mixing between the first and third generations.  At
present the constraint from $d_d^C$ is weak.  However, the further
improvement of the hadronic EDM measurements, such as the deuteron
EDM, may give a significant impact.

\begin{figure}[htb]
\begin{center}
\includegraphics[width=12pc]{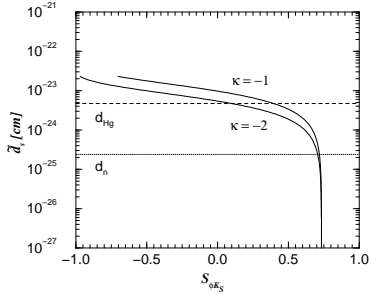} 
\caption{
The correlation between $d^C_s$ and the CP asymmetry in
$B\rightarrow\phi K_s$ ($S_{\phi K_s}$). The dashed (dotted) line is
the upperbound on $d^C_s$ from the EDM of $^{199}$Hg atom (neutron). 
See Ref.~\cite{Hisano:2004tf} for the input parameters.}
\label{fig:edm2}
\end{center}
\end{figure}

Now we discuss the implication of the hadronic EDM constraints to
other phenomena. The Belle and BaBar experiments reported recently
that the CP asymmetries in the $b$--$s$ penguin processes in the $B$
decay, including $B\rightarrow \phi K_s$, are deviated from the SM
prediction \cite{bsexp}. The $b$--$s$ penguin is the radiative process
even in the SM, and it is sensitive to beyond the SM.  It is
pointed out that the right-handed sdown mixing between the second and
third generations, which is predicted in the SUSY GUTs, may give 
sizable corrections to the processes
\cite{Moroi:2000tk}. However, the constraint on the right-handed sdown mixing 
from the hadronic EDMs implies that the deviation in the $b$--$s$
penguin processes in the SUSY  GUTs
should be suppressed
\cite{Hisano:2003iw,Hisano:2004tf}.  In Fig.~\ref{fig:edm2} we show
the correlation between $d^C_s$ and the CP asymmetry in
$B\rightarrow\phi K_s$ ($S_{\phi K_s}$) assuming the non-vanishing
right-handed sdown mixing.  From this figure, it is found that the
deviation of $S_{\phi K_s}$ from the SM prediction ($0.731\pm 0.056$)
is suppressed. In addition to this correlation, the EDMs and
$Br(\tau\rightarrow\mu\gamma)$ are also correlated with each other due
to the GUT relation among the SUSY breaking terms
\cite{Hisano:2003bd,Hisano:2004pw}. The constraint on $d_s^C$ implies
$Br(\tau\rightarrow\mu\gamma)\lsim 10^{-(7-8)}$.

\section{Leptonic EDMs}

The leptonic EDMs, such as muon and electron EDMs, are also predicted
in the SUSY GUTs. Even in the minimal SUSY seesaw model the EDMs are
generated by the threshold correction to the FVSB terms, however, they
are smaller
\cite{Ellis:2001yz}. In the SUSY GUTs, the left-handed slepton mixings 
are induced by the neutrino mixing while those for the
right-handed sleptons are parameterized by the CKM mixing at the GUT
scale. The leptonic EDMs are correlated with $Br(\mu\rightarrow e
\gamma)$, and null result in the $\mu\rightarrow e \gamma$ search
gives bounds on the leptonic EDMs.

The effective operator for $\mu\rightarrow e \gamma$ is $H=
\overline{e}(F\sigma)(A_R P_L +A_L P_R)\mu$. While $A_L$ has the various
contributions, the dominant contribution to $A_R$ comes from a diagram
proportional to $m_\tau$.  The electron (muon) EDM
is also generated by a similar diagram proportional to $m_\tau$.
Since $A_L$ and $A_R$ are not interfered with each other, the
experimental bound, $Br(\mu\rightarrow e
\gamma)<1.2\times 10^{-11}$, gives the upperbounds
on the leptonic EDMs as $|d_e|\lsim 3\times 10^{-26}
|U_{e3}/U_{\mu3}|~e~cm$ and $|d_\mu|\lsim 3\times 10^{-26}
|V_{32}/V_{31}|~e~cm$. Here, $V$ is the CKM matrix at the GUT
scale. Thus, $|d_e|\lsim 10^{-{(26-27)}}~e~cm$, and this is comparable to
the current experimental bound. For the muon EDM, $|V_{32}/V_{31}|\sim
5$ implies $|d_\mu|\lsim 10^{-25}~e~cm$. If $V_{31}$
is much smaller than the measured value at low energy and
$Br(\mu\rightarrow e
\gamma)$ is more suppressed, $V_{32}\sim 0.04$ leads to 
$|d_\mu|\lsim 10^{-24}~e~cm$ \cite{Hisano:2003bd}.

\section{Summary}
In this article, the hadronic and leptonic EDMs in the SUSY GUTs are
reviewed. The EDMs are sensitive to the FVSB for squarks
and sleptons generated by the GUT scale interaction. The EDMs are
traditional, however, still stringent constraints on the model beyond
the SM, and it is important to take correlations of the EDMs with other 
low energy observables.

\end{document}